\documentclass{article}
\usepackage[utf8]{inputenc}
\usepackage{fullpage}
\usepackage{float}
\usepackage{amsmath,amsthm, amsfonts,mathtools,mathabx}
\usepackage{graphicx,caption}
\usepackage{xcolor}
\usepackage{authblk}
\usepackage{arydshln}
\usepackage{pifont}
\usepackage{comment}
\usepackage{multirow}
\usepackage{caption}
\usepackage{subcaption}
\usepackage{cprotect}
\usepackage{hyperref}
\usepackage{authblk}
\usepackage{color, colortbl}
\usepackage{xr}
\usepackage{multirow}
\usepackage{multicol}
\makeatletter
\newcommand*{\addFileDependency}[1]{
  \typeout{(#1)}
  \@addtofilelist{#1}
  \IfFileExists{#1}{}{\typeout{No file #1.}}
}
\makeatother

\newcommand*{\myexternaldocument}[1]{%
    \externaldocument{#1}%
    \addFileDependency{#1.tex}%
    \addFileDependency{#1.aux}%
}
\myexternaldocument{supplement}

\DeclareMathOperator{\E}{\textnormal{\mbox{E}}}
\newcommand\independent{\protect\mathpalette{\protect\independenT}{\perp}}
\def\independenT#1#2{\mathrel{\rlap{$#1#2$}\mkern2mu{#1#2}}}

\let\proglang=\textsf
\newcommand{\pkg}[1]{{\fontseries{b}\selectfont #1}}

\newcommand{\keywords}[1]{\textbf{\textit{Keywords---}} #1}

\usepackage[citestyle = numeric-comp, bibstyle =  apa, sorting = none]{biblatex}
\makeatletter
\RequireBibliographyStyle{numeric}

    \title{CausalMetaR: An R package for performing causally interpretable meta-analyses}
\author[1,2$\dagger$]{Guanbo Wang\thanks{Correspondence: Guanbo Wang, CAUSALab, Harvard T.H. Chan School of Public Health, Boston, MA, 02115, U.S.A. Email: gwang@hsph.harvard.edu
$\dagger$ Equal contributed author}}
\author[3$\dagger$]{Sean McGrath}
\author[4$\dagger$]{Yi Lian}

\affil[1]{CAUSALab, Harvard T.H. Chan School of Public Health, Boston, MA, U.S.A}
\affil[2]{Department of Epidemiology, Harvard T.H. Chan School of Public Health, Boston, MA, U.S.A}
\affil[3]{Department of Biostatistics, Harvard T.H. Chan School of Public Health, Boston, MA, U.S.A}
\affil[4]{Department of Biostatistics, Epidemiology and Informatics, University of Pennsylvania, Philadelphia, PA, U.S.A}
\date{}
\begin{document}

\maketitle

\begin{abstract}
Researchers would often like to leverage data from a collection of sources (e.g., meta-analyses of randomized trials, multi-center trials, pooled analyses of observational cohorts) to estimate causal effects in a target population of interest. However, because different data sources typically represent different underlying populations, traditional meta-analytic methods may not produce causally interpretable estimates that apply to any reasonable target population. In this paper, we present the \pkg{CausalMetaR} \proglang{R} package, which implements robust and efficient methods to estimate causal effects in a given internal or external target population using multi-source data. The package includes estimators of average and subgroup treatment effects for the entire target population. To produce efficient and robust estimates of causal effects, the package implements doubly robust and non-parametric efficient estimators and supports using flexible data-adaptive (e.g., machine learning techniques) methods and cross-fitting techniques to estimate the nuisance models (e.g., the treatment model, the outcome model). We briefly review the methods, describe the key features of the package, and demonstrate how to use the package through an example. The package aims to facilitate causal analyses in the context of meta-analysis.
\end{abstract}


\section{Introduction}
We consider the setting where one would like to estimate the causal effects of an intervention in a particular target population using multi-source data. While Average Treatment Effects (ATEs) are often of interest in causal analyses, treatment effects may vary based on specific patient characteristics. In such scenarios, estimating Subgroup Treatment Effects (STEs) can help clinicians tailor treatment strategies for patients and lay the foundation for generating future hypotheses \cite{angus2021heterogeneity, dahabreh2023toward, wang2024using}.

When data from multiple sources are available, one may naturally consider pooling the data across sources to estimate such causal effects. As a primary example, meta-analyses frequently pool data across several randomized trials. Other examples include analyses of multi-center trials, pooled analyses of observational cohorts, and combined trials and observational cohorts.

Conventional meta-analytic methods pool data across sources by taking a weighted average of the source-specific estimates of the outcome of interest, where the weights are related to the precision of the sources. Typically, meta-analysis practitioners employ random effects models to account for differences in the source populations. However, the pooled estimate from such approaches generally does not have a clear causal interpretation because it does not pertain to a well-defined target population \cite{dahabreh2020towards}. 

Several methods have been recently developed to estimate causal effects for specific target populations when data are available from multiple sources. Broadly, these methods can be categorized as those that estimate causal effects in a so-called \emph{internal target population} (i.e., the population underlying one of the sources contributing covariate, treatment and outcome data to the analyses) versus an \emph{external target population} (i.e., a population underlying another source where only covariate data can be obtained from the population).
 
Specifically, Robertson et al. \cite{robertson2021center} and Dahabreh et al. \cite{dahabreh2023efficient} developed methods to estimate ATEs in internal and external target populations, respectively. These methods have subsequently been extended by Wang et al. \cite{wang2024efficient} to estimate STEs in internal and external target populations, where the subgroups are defined based on a categorical effect modifier.  

These methods can be challenging to apply for a few reasons. These methods require fitting a number of models to estimate the so-called \emph{nuisance functions}, such as the conditional outcome mean, the conditional probability of receiving treatment, and the conditional probability of the source index. To handle high-dimensional covariates and help protect against model misspecification, data analysts may want to use flexible machine methods to estimate the nuisance functions. Constructing valid confidence intervals for these estimators when using machine learning methods requires sample splitting and cross-fitting, which can be challenging to implement. Consequently, the lack of available software is one barrier to applying these methods in practice. Additionally, implementing these different methods in one software tool can help researchers easily perform secondary analyses, such as estimating treatment effects in different target populations and subgroups. 

In this article, we present the \pkg{CausalMetaR} \proglang{R} package \cite{CausalMetaR} for performing causally interpretable meta-analyses. The package can be used to estimate ATEs and STEs in internal and external target populations. Users of the package can apply a wide range of flexible models to estimate the nuisance functions. The package implements sample splitting and cross-fitting procedures, which are crucial for obtaining valid confidence intervals when using flexible models for the nuisance functions. Users can also generate forest plots of the causal effect estimates in the internal populations. The package requires individual participant data from the data sources (e.g., observational studies and/or randomized trials). The package can be downloaded from the Comprehensive R Archive Network (CRAN) at \url{https://CRAN.R-project.org/package=CausalMetaR}.

In the following section, we describe the statistical methods implemented in \pkg{CausalMetaR}. We describe how to use the package in Section \ref{sec: software} and illustrate an example application in Section \ref{sec: example}. We conclude with a discussion in Section \ref{sec: discussion}.

\section{Methods} \label{sec: methods}
In this section, we present the methods used by the package. The methods to estimate the ATEs were developed by Robertson et al. \cite{robertson2021center} (for an internal target population) and Dahabreh et al. \cite{dahabreh2023efficient} (for the external target population). The methods to estimate the STEs were developed by Wang et al.\cite{wang2024efficient} (for both types of target populations).

\subsection{Target parameters, assumptions, and identification}\label{sec: methods_ATE}
Consider a collection of datasets comprising $m$ distinct datasets, each containing $n_j, j=1,\dots,m$ random samples drawn from a unique population. The total sample size of these $m$ datasets is thus $n=n_1+\dots+n_m$. Each dataset in this collection comprises information on the outcome variable $Y$, the data source index $S\in\mathcal{S}=\{1, \dots, m\}$, the binary treatment assignment $A$ ($a\in\{0, 1\}$), and the potential high-dimensional covariate matrix $X$. We do not require the source of each data to be consistent. That is, some data can be collected from clinical trials, and some data can be collected from observational cohorts.

\begin{figure}
\centering
\begin{minipage}[t]{.45\textwidth}
  \centering
  \includegraphics[width=.5\linewidth]{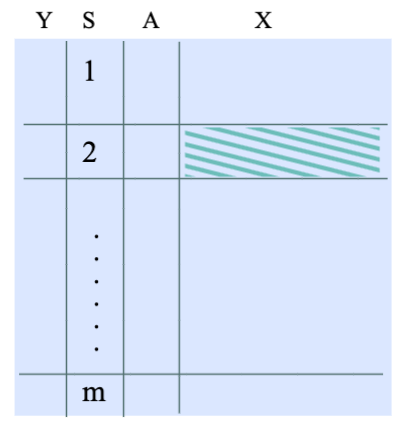}
  \captionof{figure}{The data structure required when the target population is an internal population in the multi-source data. The shaded dataset represents an example of a target population.}
  \label{fig: data structure_a}
\end{minipage}%
\hfill
\begin{minipage}[t]{.45\textwidth}
  \centering
  \includegraphics[width=.5\linewidth]{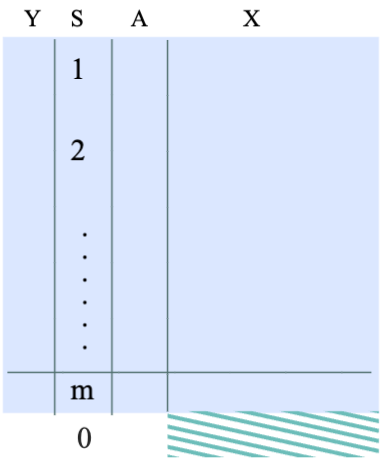}
  \captionof{figure}{The data structure required when the target population is the external target population. The shaded dataset represents the target population.}
  \label{fig: data structure_b}
\end{minipage}
\end{figure}

Throughout this paper, we use a counterfactual framework \cite{neyman1923application, rubin1974estimating, robins2000d}. Define the counterfactual outcome $Y^a$ as the outcome had the subject received treatment $a$. When the target population is the internal target population, we define the target parameter as the counterfactual means difference in the target population, denoted by $\E(Y^1 - Y^0 | S=s), \forall s \in \mathcal{S}$. See the data structures in the left panel of Figure \ref{fig: data structure_a}.

However, sometimes investigators are interested in a target population that is not represented by any of the $m$ sources of the dataset, but by an external source, in which the outcome and treatment information may be missing. In such cases, we define the target parameter as $\E(Y^1 - Y^0 | S=0)$, where $S=0$ indicates the external data. The sample size is $N=n+n_0$, where $n_0$ is the sample size of external dataset. See the data structure in the right panel of Figure \ref{fig: data structure_b}.

In both scenarios, we can estimate the STEs. Denote the interested categorical effect modifier by $\widetilde X\subset X$. When the target population is an internal target population, we define the target parameters as $\E(Y^1 - Y^0 |\widetilde X = \widetilde x, S=s), \forall s \in \mathcal{S}$, when the target population is the external population, the target parameter is $\E(Y^1 - Y^0 |\widetilde X = \widetilde x, S=0)$.

The above parameters of interest are based on counterfactual outcomes, which are not fully observed. We provide sufficient identifying conditions for the target parameters below. 

\noindent
\emph{A1. Consistency:} If $A_i = a,$ then $Y^a_i = Y_i, \forall a\in\{0, 1\}$ and $i$. 

\noindent
\emph{A2. Exchangeability over $A$:} $Y^a \independent A | (X, S=s), \forall a\in\{0, 1\}, s \in \mathcal{S}$.

\noindent
\emph{A3. Positivity of the probability of treatment:} If $f(x, S = s) \neq 0$, then $\Pr(A = a | X = x, S = s) > 0, \forall a\in\{0, 1\}, s \in \mathcal{S}$.

\noindent
\emph{A4. Exchangeability over $S$:} $\forall a\in\{0, 1\}, Y^a \independent S | X$.

It is important to emphasize that the assumptions of exchangeability and treatment positivity are assumed to hold within each individual data source, as opposed to being held to other multi-source datasets. These assumptions must be scrutinized in light of substantive knowledge. While it is not possible to empirically test assumptions \emph{A2} and \emph{A4}, they do imply a testable condition: $Y\independent S|(X, S\in \mathcal{S}, A=a)$. For a more comprehensive discussion of these assumptions and their implications, as well as potential alternative, less restrictive identifying conditions, interested readers can refer to \cite{dahabreh2023efficient, wang2024efficient, dahabreh2024learning, wang2024causal}

Through assumptions \emph{A1} to \emph{A4}, it can be shown that the counterfactual outcome means in a target population (and subgroup) are identifiable using observed data, as shown Table \ref{tab: identification}.
\begin{table}[ht]
    \centering
    \caption{Identification results for the marginal counterfactual outcomes means in a target population (and subgroup)}
    \begin{tabular}{lll}\hline
     &Counterfactual outcome mean   &  Identified quantity  \\\hline
    ATE-Internal&$\E(Y^a |S=s)$ & $\psi_{s, a}=\E\big\{\E(Y |A=a, X)| S=s\big\}$\\ 
    ATE-External&$\E(Y^a |S=0)$ & $\psi_{a}=\E\big\{\E(Y |A=a, X)| S=0\big\}$\\
    STE-Internal&$\E(Y^a |\widetilde X = \widetilde x, S=s)$ & $\phi_{s, a}(\widetilde x)=\E\big\{\E(Y |A=a, X)| \widetilde{X}=\widetilde{x}, S=s\big\}$     \\
    STE-External&$\E(Y^a |\widetilde X = \widetilde x, S=0)$ & $\phi_{a}(\widetilde x)=\E\big\{\E(Y |A=a, X)| \widetilde{X}=\widetilde{x}, S=0\big\}$     \\\hline
    \end{tabular}
    \label{tab: identification}
\end{table}
Then the target parameters can be identified by the difference of those means. For example, the STE for a internal target population $\E(Y^1-Y^0 |\widetilde X = \widetilde x, S=s)$ can be identified by $\phi_{s, 1}(\widetilde x)-\phi_{s, 0}(\widetilde x)$ (see Table \ref{tab: identification}).

\subsection{Estimation}\label{sec: Estimation}
\subsubsection{Influence function-based estimator}\label{sec: Estimator}
While one can estimate the target parameters using the above identification result by replacing $\E(Y |A=a, X)$ with corresponding model-based estimators and replacing expectations conditional on $S$ (and $\widetilde X$) with sample averages, correctly specifying parametric models for the outcome model may be challenging, especially when the covariates are high-dimensional. Incorrect model specification will result in inconsistent estimates for the target parameters. Influence functions can be used to derive non-parametric and efficient estimators of the target parameters. Specifically, one can estimate the target parameters by solving the estimating equations, which equates the empirical mean of the influence functions of the target parameters to zero. 

Such estimators can be presented as linear combinations of the following nuisance functions: 
\begin{enumerate}
    \item Outcome model $\E(Y |A=a, X)$, the outcome means conditional on a specific treatment and covariates.
    \item Source model $\Pr(S=s|X), \forall s\in\mathcal{S}$, the probability of source indices conditional on covariates in the multi-source data. 
    \item External model $\Pr(S=0|X)$, the probability of a subject is from the external population conditional on covariates (only applicable when estimating ATEs and STEs for the external target population)
    \item Propensity score model $\Pr(A=1|X, S=s), \forall s\in\mathcal{S}$, the probability of treatment conditional on covariates and source index in the multi-source data. 
\end{enumerate}

For example, the estimator for $\phi_{s, a}(\widetilde x)$ is given by 
\begin{align}
\label{eq: psi}
    \widehat \phi_{s, a}(\widetilde x)=\dfrac{1}{\sum_{i=1}^{n}M_i}\sum_{i=1}^{n}
    \Big[M_i\widehat \E(Y_i | A_i=a, X_i)
    +I(\widetilde{X}_i=\widetilde{x}_i)
    \dfrac{A_i\widehat \Pr(S_i=s|X_i)}{\widehat \Pr(A_i=a|X_i)}\big\{Y_i-\widehat \E(Y_i | A_i=a, X_i)\big\}\Big]
\end{align}
where $M_i=I(\widetilde{X}_i=\widetilde{x}_i, S_i=s)$ and hats denote estimators of the respective parameter. Then  $\E(Y^1-Y^0 |\widetilde X = \widetilde x, S=s)$ can be estimated by $\widehat \phi_{s, 1}(\widetilde x)-\widehat \phi_{s, 0}(\widetilde x)$. We compute $\widehat \Pr(A_i=a|X_i)$ by $\sum_{s\in\mathcal{S}}\widehat\Pr(A_i=1|S_i=s, X_i=x)\widehat\Pr(S_i=s|X_i=x)$. These are discussed in the following subsection. Similar estimators apply to the other target parameters.

\subsubsection{Nuisance function estimator}\label{sec: nuisance}
The \pkg{CausalMetaR} package leverages the rich capabilities of the \pkg{SuperLearner} package \cite{superlearner} to estimate the nuisance functions described in Section \ref{sec: Estimator}. Doing so gives users a wide array of flexible and robust approaches to estimate the nuisance functions under a unified user interface. Specifically, users can apply one of the following three broad approaches:
\begin{enumerate}
    \item Users can choose from a range of 41 parametric (e.g., generalized linear models) and nonparametric models (e.g., neural networks, random forests, support vector machines) to estimate the nuisance functions. This can be done by specifying a single algorithm in the library. The model hyperparameters (if applicable) can be fit based on cross-validation.
    \item Users can fit multiple models and combine them using the  SuperLearner (SL) method. Specifically, SL can either (a) choose the best performing model among the list of candidate models specified by the user, or (b) take an optimally weighted average of the candidate models. Model selection or model averaging is performed based on cross-validation.
    \item In the event that users would like to apply a model that is not implemented in \pkg{SuperLearner} (e.g., a Classification And Regression Tree (CART) model), the package allows users to specify their own custom model.
\end{enumerate}
In each of these approaches, SL supports parallel computing.

\subsubsection{Stratified sample splitting and cross-fitting}
Sample splitting and cross-fitting \cite{DML} can be used in estimation to allow users to have a broader choice of data-adaptive (e.g., machine-learning) methods to estimate the nuisance functions and guarantee efficiency \cite{zivich2021machine}. 

\begin{figure}[ht]
    \centering
    \includegraphics[width=1\textwidth]{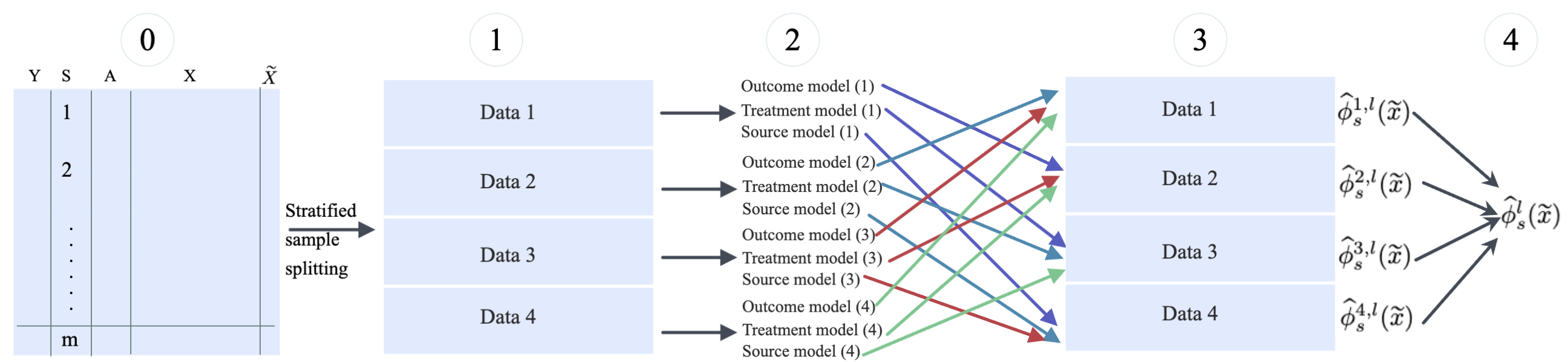}
    \caption{The cross-fitting procedure for estimating $\phi_s(\widetilde x)$ in each of the replication. 
    }
    \label{fig:CF}
\end{figure}

Broadly, this is a iterative procedure. In each replication, the data are split stratified by the data source (and subgroup when estimating STEs), then the nuisance functions and the target parameter are estimated in separate subsamples. The final estimate is the average of the estimates from each subsample. Taking estimating $\phi_s(\widetilde x)$ as an example, Figure \ref{fig:CF} illustrates this procedure in one replication. 

More specifically, the procedure implemented in \pkg{CausalMetaR} is detailed in Table \ref{table: algorithm}.

\begin{table}[ht]
    \centering
    \caption{Procedures of cross-fitting for estimating $\phi_s(\widetilde x)$}
    \begin{tabular}{cp{15cm}}\hline
    \textbf{Step} &  \textbf{Procedure}\\\hline
    0    & Data input\\
    &\quad \textbf{\textit{Repeat Steps 1-4, $L$ times, and for each replication $l$:}} (See Figure \ref{fig:CF}) \\
    1 & The data are partitioned into four approximately equal-sized sample splits, denoted by Data $k$. The partition is stratified by source index (and subgroup when estimating STEs).\\
    2 & The outcome, treatment, and source nuisance models are fit in each Data $k$. \\
    3 & The target parameter estimate $\widehat \phi_s^{k, l}(\widetilde x), \forall k=1,\dots,4$ is obtained by (\ref{eq: psi}) using Data $k$, where the nuisance models are constructed from other data split $\forall k'\neq k$. For example, use Data 1 to obtain $\widehat \phi_s^{1, l}(\widetilde x)$, where outcome, treatment, and source models are constructed by Data 2, 3, and 4, respectively.\\
    4 & The target parameter is calculated from the mean of the predictions across all splits, i.e., $\widehat \phi_s^l(\widetilde x)=\sum_{k=1}^{4}\widehat \phi_s^{k, l}(\widetilde x)/4$. The same for the variance estimation. \\
    5 & The overall point estimate is $\widehat \phi_s(\widetilde x)=\mathrm{median}\{\widehat \phi_s^l(\widetilde x)\}$, and the overall estimated variance is estimated by $\mathrm{median}[Var\{\widehat \phi_s^l(\widetilde x)\}+\{\widehat \phi_s^l(\widetilde x)-\widehat \phi_s(\widetilde x)\}^2]$, $\forall l=1,\dots,L$\\
    \hline
    \end{tabular}
    \label{table: algorithm}
\end{table} 

When the target population is the external population, we need to estimate one more nuisance function--the external model. In such cases, the data should be split into five sets in Step 1 in Figure \ref{fig:CF} and Table \ref{table: algorithm}, all other procedures are the same. Note that because Step 1 involves stratified sample splitting, the sample size of the multi-source data may not be large enough. For instance, if the data are collected from multiple early-phase trials, sample splitting and cross-fitting are not recommended. 

\subsection{Inference}

\subsubsection{Properties of estimators}
The estimators implemented in the package have the following statistical properties. The biases of the estimators depend on how the four nuisance functions described in Section \ref{sec: Estimation} are estimated. When these functions are estimated using parametric models, the estimators are consistent if either the outcome model or the collection of the other nuisance function models is correctly specified, which is known as double robustness. When the nuisance functions are estimated non-parametrically, the rates of convergence of the estimators depend on the convergence rates of the non-parametric models. Specifically, if all the nuisance function estimators converge at a rate of $n^{1/4}$ where $n$ is the sample size, estimators converge to the truth at a rate of $n^{1/2}$, which is known as rate robustness. 

Additionally, under conditions \emph{A1}-\emph{A4}, and when all the nuisance models are correctly specified, the estimators have the smallest variance among the class of doubly robust estimators. Further, the estimators asymptotically follow a normal distribution, allowing us to construct closed-form confidence intervals for them.

\subsubsection{Simultaneous confidence intervals}
In estimating STEs, when investigators are simultaneously interested in more than one subgroup, using confidence intervals to assess the uncertainty of the target parameters may be preservative. This is because multiple subgroups are being compared simultaneously using the same data. In such cases, simultaneous confidence bands (also known as uniform confidence bands), which take multiple-comparison issues into account, are recommended to be used. Our package provides the construction of simultaneous confidence bands when estimating STEs. 
\section{Software functionality} \label{sec: software}
The \pkg{CausalMetaR} package can be downloaded from CRAN and loaded in \proglang{R} as follows:
\begin{verbatim}
install.packages("CausalMetaR")
library("CausalMetaR")
\end{verbatim}

The package has four main functions: (1) \verb|ATE_internal| estimates ATEs in internal target populations, (2) \verb|ATE_external| estimates the ATE in an external target population, (3) \verb|STE_internal| estimates STEs in internal target populations, and (4) \verb|STE_external| estimates STEs in an external target population. The main arguments used in these functions are summarized in Table \ref{table:arguments}. 

The following subsections detail how to specify the data and working models for these functions as well as the output of these functions. 

\begin{table}[ht]  
\begin{center}
\caption{Summary of the arguments in the main functions in \pkg{CausalMetaR}. \label{table:arguments}} 
\begin{tabular}{lll}
\hline
\textbf{Argument} & \textbf{Description} & \textbf{Requirements}\\
  \hline
\textbf{Data} \\
\quad \verb|Y| & Outcome &  Vector \\ 
\quad \verb|S| & Source  &  Categorical vector \\ 
\quad \verb|A| & Treatment &   Binary vector\\ 
\quad \verb|X| & Covariates &  Data frame \\
\quad \verb|EM|$\textsuperscript{a}$ &  Effect modifier & Categorical vector\\
\quad \verb|X_external|$\textsuperscript{b}$ & Covariates in external population & Data frame\\ 
\quad \verb|EM_external|$\textsuperscript{c}$ & Effect modifier in external population & Categorical vector\\
\textbf{Nuisance models} \\
\quad \verb|outcome_model_args| &  Outcome model arguments & List for \verb|SuperLearner|  \\ 
\quad \verb|source_model| &  Source model specification &  \verb|"MN.glmnet"| or \verb|"MN.nnet"| \\ 
\quad \verb|source_model_args| &  Source model arguments &  List for \verb|SuperLearner| \\ 
\quad \verb|treatment_model_type| & Treatment model type &  \verb|"joint"| or \verb|"separate"| \\ 
\quad \verb|treatment_model_args| & Treatment model arguments & List for \verb|SuperLearner| \\ 
\quad \verb|external_model_args|\textsuperscript{b} &  External model arguments & List for \verb|SuperLearner| \\ 
\textbf{Cross-fitting options} \\
\quad \verb|cross_fitting| & Use of cross-fitting &  \verb|TRUE| or \verb|FALSE| \\
\quad \verb|replications| & Number of replications & Integer\\
   \hline
\end{tabular}
\caption*{
\textsuperscript{a}Only applicable for the \texttt{STE\_internal} and \texttt{STE\_external} functions \\
\textsuperscript{b}Only applicable for the \texttt{ATE\_external} and \texttt{STE\_external} functions\\
\textsuperscript{c}Only applicable for the \texttt{STE\_external} function}
\end{center}
\end{table}

\subsection{Data Specification}

The main functions in \pkg{CausalMetaR} require users to supply \proglang{R} data frames (or vectors) containing data on the outcome, source index, treatment, and covariates. Each of these data frames (or vectors) must contain $n=\sum_{s=1}^m n_s$ rows. The argument \verb|Y| specifies the outcome data, which may be a continuous or binary vector. The data source indicator is specified by the argument \verb|S|, which can be a vector of characters or integers. The argument \verb|A| specifies the treatment data, which must be binary and coded as 0/1. The argument \verb|X| specifies the baseline covariate data, which may contain continuous, binary, and/or categorical covariates.   

There are three additional requirements on the input data sets for some of the functions in \pkg{CausalMetaR}. First, when estimating STEs, \verb|EM| (and \verb|EM_external| when using \verb|STE_external|) should be provided as a categorical effect modifier of interest. Note that \verb|X| should \textit{not} include this effect modifier. That is, \verb|X| is defined slightly differently here than the $X$ defined in Section \ref{sec: methods} (i.e., $X$=\{\verb|X|, \verb|EM|\}). Second, when estimating treatment effects in an external target population, users must provide data on the baseline covariates in the target population, specified by the argument \verb|X_external|. The ordering of the columns of \verb|X_external| must be the same as that of \verb|X|.

\subsection{Nuisance models specification}

Recall that the \pkg{CausalMetaR} package generally uses SL to estimate the working models via the \pkg{SuperLearner} \proglang{R} package \cite{superlearner}. The following subsections describe how to specify the four different working models. 

\subsubsection{Outcome model}

The outcome model is specified by the argument \verb|outcome_model_args|, which is a list that supplies arguments to the SL method (specifically the \verb|SuperLearner| function in the \pkg{SuperLearner} package). Users should specify the type of model(s) considered by SL through the \verb|SL.library| component in the list. For example, setting \verb|SL.library| to \verb|"SL.glm"| includes a generalized linear model and setting \verb|SL.library| to \verb|"SL.glmnet"| includes penalized generalized linear models. Users should generally specify whether the outcome is binary or continuous through the \verb|family| component of the list, where \verb|gaussian()| indicates a continuous outcome and \verb|binomial()| indicates a binary outcome.

For example, we can specify fitting a linear regression model with a Lasso penalty in the \verb|ATE_external| function as follows\footnote{Throughout, the inclusion of \texttt{...} in the example code denotes the inclusion of all other necessary arguments to complete the call to the function.}:
\begin{verbatim}
ATE_external(..., 
        outcome_model_args = list(family = gaussian(), SL.library = "SL.glmnet"))
\end{verbatim}

\subsubsection{Source model}

Recall that the source model has a categorical outcome. Since the current version of the \pkg{SuperLearner} package (version 2.0-28) does not fully support multinomial models, the \pkg{CausalMetaR} package has the following two options for specifying the source model. If users set  \verb|source_model| to \verb|"MN.glmnet"|, the package will fit a (penalized) multinomial logistic regression model based on the \pkg{glmnet} \proglang{R} package \cite{glmnet1, glmnet2}. If users set \verb|source_model| to \verb|"MN.nnet"| which results in a multinomial log-linear model via neural networks based on the \pkg{nnet} \proglang{R} package \cite{nnet}.

\subsubsection{Treatment model}
The treatment model is specified by the arguments \verb|treatment_model_args| and \verb|treatment_model_type|. Users should supply arguments to the SL method through the argument \verb|treatment_model_type|, similar to the case of the \verb|outcome_model_args| in the outcome model. Users can specify whether to fit a treatment model separately in each data source or whether to fit a single treatment model across all sources through the argument \verb|treatment_model_type|. Setting \verb|treatment_model_type| to \verb|"separate"| results in fitting $\Pr(A=1|X, S=s)$ by regressing $A$ on $X$ in a specific data source $s$. Setting \verb|treatment_model_type| to \verb|"joint"| results in fitting a single treatment model that includes the source as a categorical predictor, i.e., regressing $A$ on $S$ and $X$ in all the data sources.

For example, we can specify fitting a joint treatment model (across all sources) based on a logistic regression model with a Lasso penalty in the \verb|ATE_external| function as follows:
\begin{verbatim}
ATE_external(..., 
        treatment_model_args = list(family = binomial(), SL.library = "SL.glmnet"), 
        treatment_model_type = "joint")
\end{verbatim}

\subsubsection{External model}

Recall that specifying an external model is needed for the \verb|ATE_external| and \verb|STE_external| functions. The argument \verb|external_model_args| specifies the external model in the same manner as the treatment and outcome models. 

\subsection{Cross-fitting specification}
To allow for the use of data-adaptive methods (e.g., machine learning techniques) to model the nuisance functions, sample splitting and cross-fitting are required in the estimation \cite{zivich2021machine}. The use of sample splitting and cross-fitting is specified by the arguments \verb|cross_fitting| and \verb|replicates|. When \verb|cross_fitting| is set to \verb|TRUE|, the package estimates the target parameters through stratified sample splitting and cross-fitting. The argument \verb|replications| specifies the number of cross-fitting replications. To achieve non-sensitive (regarding the sample splitting) results, we recommend users set \verb|replications| to \verb|100L| \cite{chernozhukov2018double}. 
Because cross-fitting can have high computational cost, we recommend users set \verb|cross_fitting| to \verb|FALSE| when users opt for parametric models to model the nuisance functions or when the sample size of multi-source data is small.

\subsection{Output}

The four main functions in \pkg{CausalMetaR} return lists which include data frames containing the potential outcome mean estimates (\verb|df_A0| and \verb|df_A1|) and treatment effect estimates (\verb|df_dif|). These data frames include the point estimates, standard error (SE) estimates, 95\% confidence intervals (CIs), and 95\% simultaneous confidence bands (SBCs) when STEs are estimated. Unlike the CIs which reflect a range of values for each (subgroup) treatment effect estimate individually, the SCBs reflect a range of values for all STE estimates simultaneously. Users can also access the fitted working models through the \verb|fit_outcome|, \verb|fit_source|, \verb|fit_treatment|, and \verb|fit_external| components of the output. Though those working models are not accessible in the output when cross-fitting is conducted, users can try fitting models without cross-fitting to check the working models first. 

The output of these functions have corresponding \verb|print| and \verb|summary| methods. Moreover, the output of the \verb|ATE_internal| and \verb|STE_internal| functions have corresponding \verb|plot| methods which generate forest plots based on the \pkg{metafor} package \cite{metafor}. These functions are illustrated in the complete example in the following section.

\section{Example} \label{sec: example}
In this section, we illustrate an application of \pkg{CausalMetaR} to an example data set in the package. 

The multi-source data set, \verb|dat_multisource|, is a data frame consisting of data from 3 sources with sample sizes of 2,312, 1,147, and 592 respectively. This data frame has columns for a continuous outcome (\verb|Y|), source indicator (\verb|S|), binary treatment (\verb|A|), effect modifier with 5 categories (\verb|EM|), and nine continuous covariates (\verb|X2|, \dots, \verb|X10|). The external data set, \verb|dat_external|, is a data frame consisting of the external data in a population with a sample size of 10,083. It has columns for the effect modifier (\verb|EM|) and the nine covariates (\verb|X2|, \dots, \verb|X10|). 

We estimate the ATE and STEs in the three internal target populations and the external target population. We use SL with (penalized) GLMs and neural networks to estimate the outcome, treatment, and external models. We use separate (source-specific) treatment models, and we use the default multinomial logistic regression model for the source model. We conducted cross-fitting with, for simplicity, 5 replications. Each of the analyses only takes a few minutes to run on a standard laptop computer (8 GB RAM, 1.1 GHz Quad-Core Intel Core i5
processor). 

We can apply the \pkg{CausalMetaR} package to perform this analysis as follows. We begin with specifying the working models:
\begin{verbatim}
outcome_model_args <- list(family = gaussian(),
                           SL.library = c("SL.glmnet", "SL.nnet", "SL.glm"))
treatment_model_args <- list(family = binomial(),
                             SL.library = c("SL.glmnet", "SL.nnet", "SL.glm"))
external_model_args = list(family = binomial(),
                           SL.library = c("SL.glmnet", "SL.nnet", "SL.glm"))
\end{verbatim}
Note that we do not explicitly specify the source model here because we will use the default source model.

The following subsections present the \proglang{R} code and output for estimating the ATEs and STEs in the internal and external target populations. In the supporting information, we present additional output from these analyses (e.g., the potential outcome means).

\subsection{Estimating the ATE in the external target population}
We apply the \verb|ATE_external| function to estimate the ATE in the external target population. Since cross-validation is used when fitting the working models (which involves random sampling), we set a random number set for reproducibility. 
\begin{verbatim}
set.seed(1234)
result_ae <- ATE_external(
  Y = dat_multisource$Y,
  S = dat_multisource$S,
  A = dat_multisource$A,
  X = dat_multisource[, 1:10],
  X_external = dat_external[, 1:10], 
  outcome_model_args = outcome_model_args,
  treatment_model_args = treatment_model_args, 
  external_model_args = external_model_args,
  cross_fitting = TRUE,
  replications = 5)
\end{verbatim}
The printed output is given below. It reports an ATE estimate of 6.63 [95\% CI: 5.94, 7.32] in the target population. 
\begin{verbatim}
AVERAGE TREATMENT EFFECT ESTIMATES IN AN EXTERNAL POPULATION

Treatment effect (mean difference) estimates:
---------------------------------------------
 Estimate     SE Lower 95% CI Upper 95% CI
   6.6294 0.1535       5.8616       7.3972
\end{verbatim}

\subsection{Estimating ATEs in the internal target populations}
We can use the \verb|ATE_internal| function to estimate the ATEs in the internal target populations as follows.
\begin{verbatim}
result_ai <- ATE_internal(
  Y = dat_multisource$Y,
  S = dat_multisource$S,
  A = dat_multisource$A,
  X = dat_multisource[, 1:10],
  outcome_model_args = outcome_model_args,
  treatment_model_args = treatment_model_args,
  cross_fitting = TRUE,
  replications = 5)
\end{verbatim}
The printed output is given below. We estimate an ATE of 6.59 [95\% CI: 6.31, 6.88] in the population of source A, 7.76 [95\% CI: 7.49, 8.03] in the population of source B, and 7.25 [95\% CI: 6.99, 7.51] in the population of source 3.
\begin{verbatim}
AVERAGE TREATMENT EFFECT ESTIMATES IN INTERNAL POPULATIONS

Treatment effect (mean difference) estimates:
---------------------------------------------
 Source Estimate     SE Lower 95% CI Upper 95% CI
      A   6.5874 0.1903       6.2145       6.9603
      B   7.7556 0.2577       7.2506       8.2606
      C   7.2916 0.3594       6.5872       7.9960
\end{verbatim}

\subsection{Estimating STEs in the external target population}
Next, we apply the \verb|STE_external| function to estimate the STEs in the external target population as follows.
\begin{verbatim}
result_se <- STE_external(
  Y = dat_multisource$Y, 
  S = dat_multisource$S, 
  A = dat_multisource$A, 
  X = dat_multisource[, 2:10], 
  EM = dat_multisource$EM,
  X_external = dat_external[, 2:10], 
  EM_external = dat_external$EM, 
  outcome_model_args = outcome_model_args, 
  treatment_model_args = treatment_model_args, 
  external_model_args = external_model_args,
  cross_fitting = TRUE,
  replications = 5)
\end{verbatim}
The printed output is given below. The STE estimates range from 5.41 [95\% CI: 4.83, 6.00] (in subgroup e) to 7.58 [95\% CI: 6.99, 8.17] (in subgroup c). 
\begin{verbatim}
SUBGROUP TREATMENT EFFECT ESTIMATES IN AN EXTERNAL POPULATION

Treatment effect (mean difference) estimates:
---------------------------------------------
 Subgroup Estimate     SE Lower 95% CI Upper 95% CI Lower 95% SCB Upper 95% SCB
        a   7.0787 0.3563       5.9088       8.2485        5.5453        8.6121
        b   5.5207 0.2321       4.5764       6.4650        4.2830        6.7585
        c   7.5709 0.1805       6.7382       8.4037        6.4794        8.6625
        d   6.5748 0.2253       5.6446       7.5051        5.3556        7.7941
        e   5.3741 0.3382       4.2343       6.5139        3.8802        6.8681
\end{verbatim}

\subsection{Estimating STEs in the internal target populations}
Last, we estimate the STEs in the internal target populations by using the \verb|STE_internal| function.
\begin{verbatim}
result_si <- STE_internal(
  Y = dat_multisource$Y, 
  S = dat_multisource$S, 
  A = dat_multisource$A, 
  X = dat_multisource[, 2:10], 
  EM = dat_multisource$EM,
  outcome_model_args = outcome_model_args, 
  treatment_model_args = treatment_model_args,
  cross_fitting = TRUE,
  replications = 5)
\end{verbatim}
The printed output for the STE estimates is given below.  
\begin{verbatim}
SUBGROUP TREATMENT EFFECT ESTIMATES IN INTERNAL POPULATIONS

Treatment effect (mean difference) estimates:
---------------------------------------------
 Source Subgroup Estimate     SE Lower 95% CI Upper 95% CI Lower 95% SCB Upper 95% SCB
      A        a   6.9197 0.5001       5.9395       7.8999        5.6345        8.2050
               b   5.4340 0.3681       4.7126       6.1555        4.4880        6.3801
               c   7.5452 0.3097       6.9383       8.1522        6.7493        8.3411
               d   6.5053 0.3630       5.7939       7.2168        5.5724        7.4382
               e   5.4595 0.5215       4.4373       6.4816        4.1192        6.7998
      B        a   8.2134 0.7554       6.7328       9.6939        6.2720       10.1547
               b   6.8396 0.5381       5.7850       7.8942        5.4567        8.2225
               c   8.7995 0.4232       7.9699       9.6290        7.7118        9.8872
               d   7.7098 0.4529       6.8223       8.5974        6.5460        8.8737
               e   6.4405 0.6975       5.0734       7.8076        4.6479        8.2332
      C        a   8.0544 1.2049       5.6929      10.4159        4.9579       11.1509
               b   6.2151 0.6711       4.8998       7.5304        4.4904        7.9398
               c   8.2620 0.6426       7.0026       9.5215        6.6106        9.9135
               d   7.1427 0.6538       5.8612       8.4242        5.4624        8.8231
               e   6.0910 0.8718       4.3823       7.7997        3.8504        8.3316
\end{verbatim}

\begin{figure} [ht]
    \centering
    \includegraphics[width=0.9\textwidth]{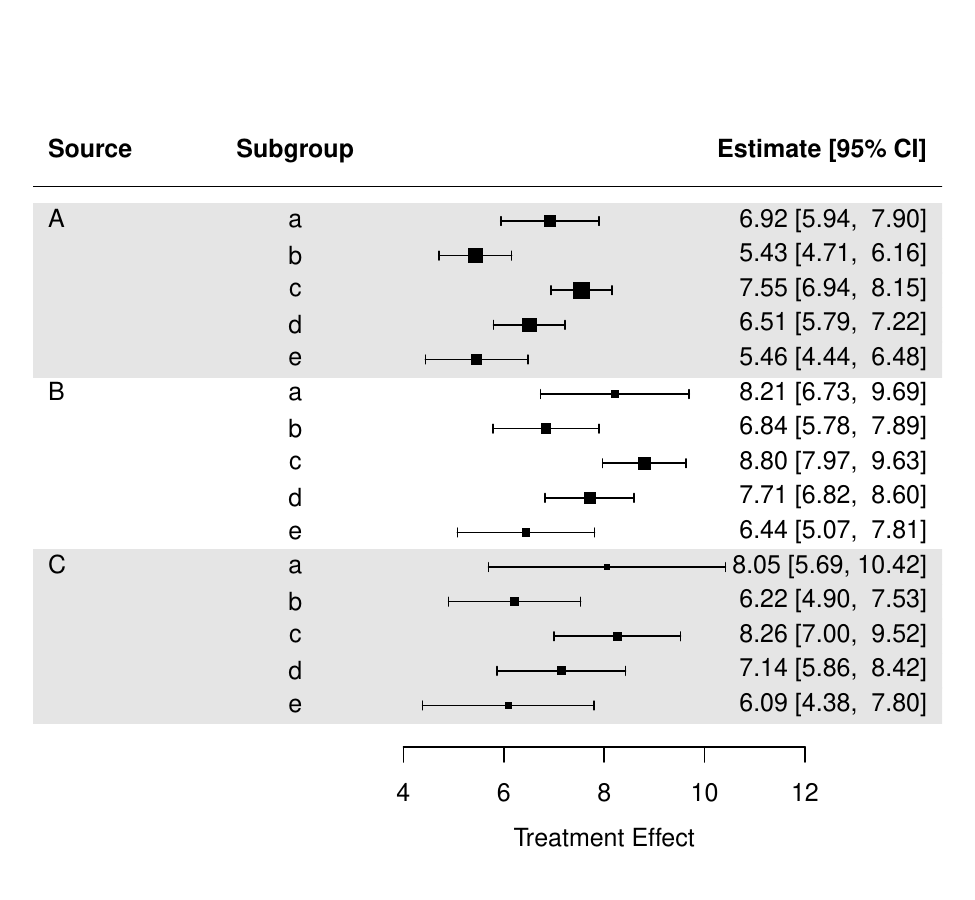}
    \caption{Forest plot of the STEs in the internal target populations. \label{fig:forest}}
\end{figure}

To generate a forest plot of the STEs in the internal target populations, we can run the command \verb|plot(result_si)| which produces Figure \ref{fig:forest}. Users can set the argument \verb|use_scb| to \verb|TRUE| to illustrate SCBs instead of CIs in the forest plot. Further options for customizing the forest plot are detailed in the package help file for \verb|plot.STE_internal|.

\section{Discussion} \label{sec: discussion}
The \pkg{CausalMetaR} package implements state-of-science approaches to estimate ATEs and STEs in internal and external target populations from multi-source data. The estimators implemented in the package are (i) doubly robust in the sense that the estimators can still be consistent even if some of the nuisance function estimators are not correctly specified, (ii) efficient in the sense that they attain the smallest possible asymptotic variance under some mild regularity conditions, and (iii) are asymptotically normally distributed which allows for constructing valid confidence intervals. Users of the package can employ a number of different modelling approaches to estimate the nuisance functions, ranging from parametric approaches such as generalized linear models to highly flexible machine learning approaches such as neural networks. 

The package implements the methods developed by Dahabreh et al. \cite{dahabreh2020towards}, Robertson et al. \cite{robertson2021center}, and Wang et al.\cite{wang2024efficient}. Such causally-interpretable meta-analyses can be easily adapted to solve many real-world problems, such as estimating causal effects for a study in a meta-analysis and estimating causal effects for a trial when combining data collected from other studies \cite{wang2023evaluating, ung2024combining, wang2024extending}. Other approaches for performing causally interpretable meta-analyses have been proposed. Vo et al. \cite{vo2019novel, vo2021assessing} developed estimators of average treatment effects in internal source populations for individual participant data (IPD) meta-analyses with case-mix heterogeneity. However, these approaches are not doubly robust and do not consider external target populations. Moreover, IPD network meta-analysis approaches have been developed to estimate a global treatment importance metric and heterogeneous treatment effects in the entire multi-source population \cite{wang2020estimating,liu2022modeling,siddique2019causal}. These approaches are not directly comparable to those implemented in \pkg{CausalMetaR} because they do not pertain to the same target populations. 

In this work, users must designate a categorical effect modifier of interest. This applies to cases where the users have strong a priori knowledge of which variables are the effect modifiers and are interested in certain of them. When the effect modifiers are unknown, one can use data-driven methods to explore the effect modifier and proceed with the analysis \cite{bargagli2020causal,ngufor2023identifying,jaman2024penalized}. Furthermore, The assumption A4 is untestable and may be too strong in some scenarios. Such an assumption can be replaced by weaker ones, such as the transportability assumption of relative effect measures \cite{dahabreh2024learning,wang2024causal}. In addition, when practitioners have prior information about the covariates (such as a variable is an interaction of other main terms), one can integrate this information into the nuisance models. Such integration would potentially improve the estimation accuracy \cite{wang2021general,bouchard2022predictive,wang2022structured,wang2024structured}.

While the current version of \pkg{CausalMetaR} (i.e., version 0.1.2) is stable and complete, we intend to add several features in future versions of the package. In particular, we aim to include methods that estimate STEs in internal and external target populations when the effect modifier ($\widetilde X$) is continuous \cite{kunzel2019metalearners,wang2024continuoustime}. Additionally, we would like to include methods that estimate quantile treatment effects in internal and external target populations \cite{firpo2007efficient,sun2021causal}. However, methodological work establishing such methods is needed first.

\section*{Authours contribution}
GW, SM, YL: Conceptualization, Methodology, Investigation, Software, Validation, Visualization, Project administration, Drafting. All authors approved the final version for publication.

\section*{Acknowledgements}
Dr. Issa J. Dahabreh has made significant contributions to this work, particularly in methods development, software design and development, and manuscript writing, and intends to be included as an author in the next revision. Due to extenuating circumstances, he was unable to participate in the current revision process and is, therefore, not listed as an author in the current draft.

GW was supported by the Patient-Centered Outcomes Research Institute (PCORI) awards ME-2021C2-22365. SM was supported by the National Science Foundation Graduate Research Fellowship Program under Grant No.\ DGE2140743. Any opinions, findings, conclusions, or recommendations expressed in this material are those of the authors and do not necessarily reflect the views of the above funding agencies.

\section*{Supporting Information}
Additional supporting information can be found online in the Supporting Information section at the end of this article.

\section*{Conflict of interest}
The authors declare no potential conflict of interest.

\section*{Data/code availability statement}
The source code of the software and data presented in this paper are publicly available on GitHub (\url{https://github.com/ly129/CausalMetaR}).

\section*{Highlights}
\textbf{What is already known}
\begin{itemize}
    \item Several methods have been recently developed to estimate causal effects in well-defined target populations from multi-source data (e.g., primary studies in a meta-analysis).
    \item There are no existing software tools implementing these methods.
\end{itemize}
\textbf{What is new}
\begin{itemize}
    \item We present the free, open-source \pkg{CausalMetaR} package for estimating causal effects in a well-defined target population from multi-source data.
    \item \pkg{CausalMetaR} facilitates estimating average and subgroup treatment effects in internal target populations and external target populations.
    \item \pkg{CausalMetaR} allows users to use machine-learning methods and cross-fitting to estimate the target parameters, and the estimators are doubly/rate robust, non-parametrically efficient, and asymptotically normal.
\end{itemize}
\textbf{Potential impact for Research Synthesis Methods readers}
\begin{itemize}
    \item \pkg{CausalMetaR} can help evidence synthesis practitioners perform more rigorous causal analyses
\end{itemize}
\printbibliography

@article{dahabreh2023efficient,
  title={Efficient and robust methods for causally interpretable meta-analysis: Transporting inferences from multiple randomized trials to a target population},
  author={Dahabreh, Issa J and Robertson, Sarah E and Petito, Lucia C and Hern{\'a}n, Miguel A and Steingrimsson, Jon A},
  journal={Biometrics},
  year={2023},
  publisher={Wiley Online Library}
}

@article{wang2024continuoustime,
      title={Continuous-time structural failure time model for intermittent treatment}, 
      author={Guanbo Wang and Siyi Liu and Shu Yang},
      year={2024},
      journal={arXiv preprint arXiv:2401.15806},
      url={https://arxiv.org/abs/2401.15806}
}

@article{jaman2024penalized,
      title={Penalized G-estimation for effect modifier selection in the structural nested mean models for repeated outcomes}, 
      author={Ajmery Jaman and Guanbo Wang and Ashkan Ertefaie and Michèle Bally and Renée Lévesque and Robert Platt and Mireille Schnitzer},
      year={2024},
journal={arXiv preprint arXiv:2402.00154},
      url={https://arxiv.org/abs/2402.00154}
}

@article{wang2024efficient,
      title={Efficient estimation of subgroup treatment effects using multi-source data}, 
      author={Guanbo Wang and Alexander Levis and Jon Steingrimsson and Issa Dahabreh},
      year={2024},
      journal={arXiv preprint arXiv:2402.02684},
      url={https://arxiv.org/abs/2402.02684}
}

@article{wang2024causal,
      title={Causal inference under transportability assumptions for conditional relative effect measures}, 
      author={Guanbo Wang and Alexander Levis and Jon Steingrimsson and Issa Dahabreh},
      year={2024},
      journal={arXiv preprint arXiv:2402.02702},
      url={https://arxiv.org/abs/2402.02702}
}

@article{dahabreh2020towards,
  title={Towards causally interpretable meta-analysis: transporting inferences from multiple randomized trials to a new target population},
  author={Dahabreh, Issa J and Petito, Lucia C and Robertson, Sarah E and Hern{\'a}n, Miguel A and Steingrimsson, Jon A},
  journal={Epidemiology (Cambridge, Mass.)},
  volume={31},
  number={3},
  pages={334},
  year={2020},
  publisher={NIH Public Access}
}

@Manual{superlearner,
    title = {SuperLearner: Super Learner Prediction},
    author = {Eric Polley and Erin LeDell and Chris Kennedy and Mark {van der Laan}},
    year = {2021},
    note = {R package version 2.0-28},
    url = {https://CRAN.R-project.org/package=SuperLearner},
  }

@article{vo2019novel,
  title={A novel approach for identifying and addressing case-mix heterogeneity in individual participant data meta-analysis},
  author={Vo, Tat-Thang and Porcher, Raphael and Chaimani, Anna and Vansteelandt, Stijn},
  journal={Research synthesis methods},
  volume={10},
  number={4},
  pages={582--596},
  year={2019},
  publisher={Wiley Online Library}
}

@article{vo2021assessing,
  title={Assessing the impact of case-mix heterogeneity in individual participant data meta-analysis: Novel use of I 2 statistic and prediction interval},
  author={Vo, Tat-Thang and Porcher, Rapha{\"e}l and Vansteelandt, Stijn},
  journal={Research Methods in Medicine \& Health Sciences},
  volume={2},
  number={1},
  pages={12--30},
  year={2021},
  publisher={SAGE Publications Sage UK: London, England}
}

@Article{metafor,
    title = {Conducting meta-analyses in {R} with the {metafor} package},
    author = {Wolfgang Viechtbauer},
    journal = {Journal of Statistical Software},
    year = {2010},
    volume = {36},
    number = {3},
    pages = {1--48},
    doi = {10.18637/jss.v036.i03},
  }

@article{robertson2021center,
  title={Center-specific causal inference with multicenter trials: reinterpreting trial evidence in the context of each participating center},
  author={Robertson, Sarah E and Steingrimsson, Jon A and Joyce, Nina R and Stuart, Elizabeth A and Dahabreh, Issa J},
  journal={arXiv preprint arXiv:2104.05905},
  year={2021}
}

@Article{glmnet1,
    title = {Regularization Paths for Generalized Linear Models via Coordinate Descent},
    author = {Jerome Friedman and Robert Tibshirani and Trevor Hastie},
    journal = {Journal of Statistical Software},
    year = {2010},
    volume = {33},
    number = {1},
    pages = {1--22},
    doi = {10.18637/jss.v033.i01},
  }

@Article{glmnet2,
    title = {Elastic Net Regularization Paths for All Generalized Linear Models},
    author = {J. Kenneth Tay and Balasubramanian Narasimhan and Trevor Hastie},
    journal = {Journal of Statistical Software},
    year = {2023},
    volume = {106},
    number = {1},
    pages = {1--31},
    doi = {10.18637/jss.v106.i01},
  }

@Book{nnet,
    title = {Modern Applied Statistics with S},
    author = {W. N. Venables and B. D. Ripley},
    publisher = {Springer},
    edition = {Fourth},
    address = {New York},
    year = {2002},
    note = {ISBN 0-387-95457-0},
    url = {https://www.stats.ox.ac.uk/pub/MASS4/},
  }

@article{wang2020estimating,
  title={Estimating treatment importance in multidrug-resistant tuberculosis using Targeted Learning: An observational individual patient data network meta-analysis},
  author={Wang, Guanbo and Schnitzer, Mireille E and Menzies, Dick and Viiklepp, Piret and Holtz, Timothy H and Benedetti, Andrea},
  journal={Biometrics},
  volume={76},
  number={3},
  pages={1007--1016},
  year={2020},
  publisher={Wiley Online Library}
}

@article{liu2022modeling,
  title={Modeling treatment effect modification in multidrug-resistant tuberculosis in an individual patient data meta-analysis},
  author={Liu, Yan and Schnitzer, Mireille E and Wang, Guanbo and Kennedy, Edward and Viiklepp, Piret and Vargas, Mario H and Sotgiu, Giovanni and Menzies, Dick and Benedetti, Andrea},
  journal={Statistical methods in medical research},
  volume={31},
  number={4},
  pages={689--705},
  year={2022},
  publisher={SAGE Publications Sage UK: London, England}
}

@article{siddique2019causal,
  title={Causal inference with multiple concurrent medications: A comparison of methods and an application in multidrug-resistant tuberculosis},
  author={Siddique, Arman Alam and Schnitzer, Mireille E and Bahamyirou, Asma and Wang, Guanbo and Holtz, Timothy H and Migliori, Giovanni B and Sotgiu, Giovanni and Gandhi, Neel R and Vargas, Mario H and Menzies, Dick and others},
  journal={Statistical methods in medical research},
  volume={28},
  number={12},
  pages={3534--3549},
  year={2019},
  publisher={SAGE Publications Sage UK: London, England}
}

@article{firpo2007efficient,
  title={Efficient semiparametric estimation of quantile treatment effects},
  author={Firpo, Sergio},
  journal={Econometrica},
  volume={75},
  number={1},
  pages={259--276},
  year={2007},
  publisher={Wiley Online Library}
}

@article{sun2021causal,
  title={Causal inference for quantile treatment effects},
  author={Sun, Shuo and Moodie, Erica EM and Ne{\v{s}}lehov{\'a}, Johanna G},
  journal={Environmetrics},
  volume={32},
  number={4},
  pages={e2668},
  year={2021},
  publisher={Wiley Online Library}
}

@article{kunzel2019metalearners,
  title={Metalearners for estimating heterogeneous treatment effects using machine learning},
  author={K{\"u}nzel, S{\"o}ren R and Sekhon, Jasjeet S and Bickel, Peter J and Yu, Bin},
  journal={Proceedings of the national academy of sciences},
  volume={116},
  number={10},
  pages={4156--4165},
  year={2019},
  publisher={National Acad Sciences}
}

@article{bargagli2020causal,
  title={Causal Rule Ensemble: Interpretable Discovery and Inference of Heterogeneous Causal Effects},
  author={Bargagli-Stoffi, Falco J and Cadei, Riccardo and Lee, Kwonsang and Dominici, Francesca},
  journal={arXiv preprint arXiv:2009.09036},
  year={2020}
}

@article{bouchard2022predictive,
  title={Predictive Factors of Detectable Viral Load in HIV-Infected Patients},
  author={Bouchard, Audrey and Bourdeau, Fran{\c{c}}ois and Roger, Julien and Taillefer, Vincent-Thierry and Sheehan, Nancy L and Schnitzer, Mireille and Wang, Guanbo and Jean Baptiste Fran{\c{c}}ois, Imma Judy and Therrien, Rachel},
  journal={AIDS Research and Human Retroviruses},
  volume={38},
  number={7},
  pages={552--560},
  year={2022},
  publisher={Mary Ann Liebert, Inc., publishers 140 Huguenot Street, 3rd Floor New~…}
}

@article{wang2021general,
title={A general framework for formulating structured variable selection},
author={Guanbo Wang and Mireille Schnitzer and Tom Chen and Rui Wang and Robert W Platt},
journal={Transactions on Machine Learning Research},
issn={2835-8856},
year={2024},
url={https://openreview.net/forum?id=cvOpIhQQMN},
note={}
}

@article{wang2022structured,
  title={Integrating complex selection rules into the latent overlapping group Lasso for constructing coherent prediction models},
  author={Wang, Guanbo and Perreault, Sylvie and Platt, Robert W and Wang, Rui and Dorais, Marc and Schnitzer, Mireille E},
  journal={arXiv preprint arXiv:2206.05337},
  url={https://arxiv.org/abs/2206.05337},
  year={2024}
}

@article{ngufor2023identifying,
  title={Identifying treatment heterogeneity in atrial fibrillation using a novel causal machine learning method},
  author={Ngufor, Che and Yao, Xiaoxi and Inselman, Jonathan W and Ross, Joseph S and Dhruva, Sanket S and Graham, David J and Lee, Joo-Yeon and Siontis, Konstantinos C and Desai, Nihar R and Polley, Eric and others},
  journal={American heart journal},
  volume={260},
  pages={124--140},
  year={2023},
  publisher={Elsevier}
}

@article{zivich2021machine,
  title={Machine learning for causal inference: on the use of cross-fit estimators},
  author={Zivich, Paul N and Breskin, Alexander},
  journal={Epidemiology (Cambridge, Mass.)},
  volume={32},
  number={3},
  pages={393},
  year={2021},
  publisher={NIH Public Access}
}

@Article{robins2000d,
  author        = {Robins, James M and Greenland, Sander},
  title         = {Causal inference without counterfactuals: comment},
  journal       = {Journal of the American Statistical Association},
  year          = {2000},
  volume        = {95},
  number        = {450},
  pages         = {431--435},
  __markedentry = {[issa:]},
  publisher     = {JSTOR},
}

@article{wang2024structured,
  title={Structured learning in time-dependent Cox models},
  author={Wang, Guanbo and Lian, Yi and Yang, Archer Y and Platt, Robert W and Wang, Rui and Perreault, Sylvie and Dorais, Marc and Schnitzer, Mireille E},
  journal={Statistics in Medicine},
  year={2024},
  publisher={Wiley Online Library},
    doi = {https://doi.org/10.1002/sim.10116},
    url = {https://onlinelibrary.wiley.com/doi/abs/10.1002/sim.10116},
    eprint = {https://onlinelibrary.wiley.com/doi/pdf/10.1002/sim.10116},
}

@article{chernozhukov2018double,
  title={Double/debiased machine learning for treatment and structural parameters},
  author={Chernozhukov, Victor and Chetverikov, Denis and Demirer, Mert and Duflo, Esther and Hansen, Christian and Newey, Whitney and Robins, James},
  journal={The Econometrics Journal},
  volume={21},
  number={1},
  pages={C1--C68},
  year={2018},
  publisher={Wiley Online Library}
}

@article{DML,
    author = {Chernozhukov, Victor and Chetverikov, Denis and Demirer, Mert and Duflo, Esther and Hansen, Christian and Newey, Whitney and Robins, James},
    title = "{Double/debiased machine learning for treatment and structural parameters}",
    journal = {The Econometrics Journal},
    volume = {21},
    number = {1},
    pages = {C1-C68},
    year = {2018},
    month = {01},
    issn = {1368-4221},
    doi = {10.1111/ectj.12097},
    url = {https://doi.org/10.1111/ectj.12097},
    eprint = {https://academic.oup.com/ectj/article-pdf/21/1/C1/27684918/ectj00c1.pdf},
}

@Manual{CausalMetaR,
    title = {CausalMetaR: Causally Interpretable Meta-Analysis},
    author = {Yi Lian and Guanbo Wang and Sean McGrath and Issa Dahabreh},
    year = {2024},
    note = {R package version 0.1.2},
    url = {https://CRAN.R-project.org/package=CausalMetaR},
  }

@article{dahabreh2023toward,
  title={Toward personalizing care: assessing heterogeneity of treatment effects in randomized trials},
  author={Dahabreh, Issa J and Kazi, Dhruv S},
  journal={JAMA},
  volume={329},
  number={13},
  pages={1063--1065},
  year={2023},
  publisher={American Medical Association}
}

@article{wang2024using,
  title={Using effect scores to characterize heterogeneity of treatment effects},
  author={Wang, Guanbo and Heagerty, Patrick J and Dahabreh, Issa J},
  journal={JAMA},
  volume={331},
  number={14},
  pages={1225--1226},
  year={2024},
  publisher={American Medical Association}
}

@article{angus2021heterogeneity,
  title={Heterogeneity of treatment effect: estimating how the effects of interventions vary across individuals},
  author={Angus, Derek C and Chang, Chung-Chou H},
  journal={JAMA},
  volume={326},
  number={22},
  pages={2312--2313},
  year={2021},
  publisher={American Medical Association}
}

@article{neyman1923application,
  title={On the application of probability theory to agricultural experiments. Essay on principles},
  author={Neyman, Jerzy},
  journal={Ann. Agricultural Sciences},
  pages={1--51},
  year={1923}
}

@article{rubin1974estimating,
  title={Estimating causal effects of treatments in randomized and nonrandomized studies.},
  author={Rubin, Donald B},
  journal={Journal of educational Psychology},
  volume={66},
  number={5},
  pages={688},
  year={1974},
  publisher={American Psychological Association}
}

@article{ung2024combining,
      title={Combining an experimental study with external data: study designs and identification strategies}, 
      author={Lawson Ung and Guanbo Wang and Sebastien Haneuse and Miguel A. Hernan and Issa J. Dahabreh},
      year={2024},
      eprint={2406.03302},
      archivePrefix={arXiv},
      primaryClass={stat.ME},
      url={https://arxiv.org/abs/2406.03302v1}
}

@article{dahabreh2024learning,
  title={Learning about treatment effects in a new target population under transportability assumptions for relative effect measures},
  author={Dahabreh, Issa J and Robertson, Sarah E and Steingrimsson, Jon A},
  journal={European Journal of Epidemiology},
  pages={1--9},
  year={2024},
  publisher={Springer}
}

@article{wang2024extending,
  title={Extending Inferences from Randomized Clinical Trials to Target Populations Using Observational Data: A Scoping Review of Transportability Methods},
  author={Wang, Guanbo and Liao, Ting-Wei Ernie and Furfaro, David and Celi, Leo Anthony and Ma, Kevin Sheng-Kai},
  journal={arXiv preprint arXiv:2402.07236},
  year={2024},
  url={https://arxiv.org/abs/2402.07236}
}

\clearpage
\begin{center}
    {\Large\bfseries Supporting Information}
\end{center}
\section*{Additional results from the example} \label{sec: appendix analysis}
In this section, we present additional results from the analyses described in Section \ref{sec: example} of the main text. Specifically, we illustrate the \verb|summary| methods corresponding to the \verb|ATE_internal|, \verb|ATE_external|, \verb|STE_internal|, and \verb|STE_external| functions. In addition to the estimates of the treatment effects (i.e., the difference of potential outcome means), the \verb|summary| methods include the estimates of the potential outcome means under $A=0$ and $A=1$. 

Recall from Section \ref{sec: example} in the main text that the outputs of the \verb|ATE_external|, \verb|ATE_internal|, \verb|STE_external|, and \verb|STE_internal| functions were saved as objects named \verb|result_ae|, \verb|result_ai|, \verb|result_se|, and \verb|result_si|, respectively. The following subsections illustrate applying the \verb|summary| function to these objects.

\subsection*{Estimating the ATE in the external target population}

Running \verb|summary(result_ae)| in the R console prints the following output:

\begin{verbatim}
AVERAGE TREATMENT EFFECT ESTIMATES IN AN EXTERNAL POPULATION

Treatment effect (mean difference) estimates:
---------------------------------------------
 Estimate     SE Lower 95% CI Upper 95% CI
   6.6294 0.1535       5.8616       7.3972


Potential outcome mean estimates under A = 0:
---------------------------------------------
 Estimate     SE Lower 95% CI Upper 95% CI
  19.2657 0.0934      18.6665      19.8648


Potential outcome mean estimates under A = 1:
---------------------------------------------
 Estimate     SE Lower 95% CI Upper 95% CI
  25.8961 0.1214      25.2132      26.5790


SuperLearner libraries used:
----------------------------
Outcome model: SL.glmnet, SL.nnet, SL.glm
Treatment model: SL.glmnet, SL.nnet, SL.glm
Source model: NA (model fit via MN.glmnet)
External model: SL.glmnet, SL.nnet, SL.glm
\end{verbatim}

\subsection*{Estimating ATEs in the internal target populations}

Running \verb|summary(result_ai)| in the R console prints the following output:

\begin{verbatim}
AVERAGE TREATMENT EFFECT ESTIMATES IN INTERNAL POPULATIONS

Treatment effect (mean difference) estimates:
---------------------------------------------
 Source Estimate     SE Lower 95% CI Upper 95% CI
      A   6.5874 0.1903       6.2145       6.9603
      B   7.7556 0.2577       7.2506       8.2606
      C   7.2916 0.3594       6.5872       7.9960


Potential outcome mean estimates under A = 0:
---------------------------------------------
 Source Estimate     SE Lower 95% CI Upper 95% CI
      A  19.2107 0.1072      19.0005      19.4209
      B  20.8918 0.1457      20.6062      21.1774
      C  20.2309 0.2191      19.8014      20.6603


Potential outcome mean estimates under A = 1:
---------------------------------------------
 Source Estimate     SE Lower 95% CI Upper 95% CI
      A  25.8043 0.1576      25.4955      26.1131
      B  28.6467 0.2121      28.2310      29.0625
      C  27.5024 0.2784      26.9567      28.0482


SuperLearner libraries used:
----------------------------
Outcome model: SL.glmnet, SL.nnet, SL.glm
Treatment model: SL.glmnet, SL.nnet, SL.glm
Source model: NA (model fit via MN.glmnet)
\end{verbatim}

\subsection*{Estimating STEs in the external target population}

Running \verb|summary(result_se)| in the R console prints the following output:
\begin{verbatim}
SUBGROUP TREATMENT EFFECT ESTIMATES IN AN EXTERNAL POPULATION

Treatment effect (mean difference) estimates:
---------------------------------------------
 Subgroup Estimate     SE Lower 95% CI Upper 95% CI Lower 95% SCB Upper 95% SCB
        a   7.0787 0.3563       5.9088       8.2485        5.5453        8.6121
        b   5.5207 0.2321       4.5764       6.4650        4.2830        6.7585
        c   7.5709 0.1805       6.7382       8.4037        6.4794        8.6625
        d   6.5748 0.2253       5.6446       7.5051        5.3556        7.7941
        e   5.3741 0.3382       4.2343       6.5139        3.8802        6.8681


Potential outcome mean estimates under A = 0:
---------------------------------------------
 Subgroup Estimate     SE Lower 95% CI Upper 95% CI Lower 95% SCB Upper 95% SCB
        a  17.3465 0.2519      16.3627      18.3302       16.0571       18.6359
        b  18.3945 0.1641      17.6004      19.1886       17.3537       19.4353
        c  19.2442 0.1277      18.5439      19.9444       18.3263       20.1620
        d  20.3001 0.1593      19.5179      21.0823       19.2748       21.3254
        e  21.2464 0.2391      20.2879      22.2048       19.9901       22.5026


Potential outcome mean estimates under A = 1:
---------------------------------------------
 Subgroup Estimate     SE Lower 95% CI Upper 95% CI Lower 95% SCB Upper 95% SCB
        a  24.4198 0.2972      23.3514      25.4882       23.0194       25.8202
        b  23.9372 0.2205      23.0169      24.8575       22.7310       25.1435
        c  26.8304 0.1749      26.0108      27.6501       25.7561       27.9048
        d  26.8700 0.2141      25.9631      27.7769       25.6813       28.0587
        e  26.6380 0.3242      25.5219      27.7540       25.1751       28.1008


SuperLearner libraries used:
----------------------------
Outcome model: SL.glmnet, SL.nnet, SL.glm
Treatment model: SL.glmnet, SL.nnet, SL.glm
Source model: NA (model fit via MN.glmnet)
External model: SL.glmnet, SL.nnet, SL.glm
\end{verbatim}

\subsection*{Estimating STEs in the internal target populations}

Running \verb|summary(result_si)| in the R console prints the following output:
\begin{verbatim}
SUBGROUP TREATMENT EFFECT ESTIMATES IN INTERNAL POPULATIONS

Treatment effect (mean difference) estimates:
---------------------------------------------
 Source Subgroup Estimate     SE Lower 95% CI Upper 95% CI Lower 95% SCB Upper 95% SCB
      A        a   6.9197 0.5001       5.9395       7.8999        5.6345        8.2050
               b   5.4340 0.3681       4.7126       6.1555        4.4880        6.3801
               c   7.5452 0.3097       6.9383       8.1522        6.7493        8.3411
               d   6.5053 0.3630       5.7939       7.2168        5.5724        7.4382
               e   5.4595 0.5215       4.4373       6.4816        4.1192        6.7998
      B        a   8.2134 0.7554       6.7328       9.6939        6.2720       10.1547
               b   6.8396 0.5381       5.7850       7.8942        5.4567        8.2225
               c   8.7995 0.4232       7.9699       9.6290        7.7118        9.8872
               d   7.7098 0.4529       6.8223       8.5974        6.5460        8.8737
               e   6.4405 0.6975       5.0734       7.8076        4.6479        8.2332
      C        a   8.0544 1.2049       5.6929      10.4159        4.9579       11.1509
               b   6.2151 0.6711       4.8998       7.5304        4.4904        7.9398
               c   8.2620 0.6426       7.0026       9.5215        6.6106        9.9135
               d   7.1427 0.6538       5.8612       8.4242        5.4624        8.8231
               e   6.0910 0.8718       4.3823       7.7997        3.8504        8.3316


Potential outcome mean estimates under A = 0:
---------------------------------------------
 Source Subgroup Estimate     SE Lower 95% CI Upper 95% CI Lower 95% SCB Upper 95% SCB
      A        a  17.4385 0.2571      16.9346      17.9425       16.7777       18.0993
               b  18.1922 0.1923      17.8152      18.5691       17.6979       18.6864
               c  19.1656 0.1634      18.8454      19.4859       18.7457       19.5856
               d  20.1789 0.1905      19.8056      20.5522       19.6894       20.6684
               e  21.2993 0.2669      20.7762      21.8224       20.6133       21.9853
      B        a  18.7137 0.3794      17.9701      19.4573       17.7387       19.6888
               b  19.7522 0.2782      19.2070      20.2974       19.0373       20.4671
               c  20.8026 0.2231      20.3653      21.2400       20.2291       21.3761
               d  21.6868 0.2306      21.2348      22.1388       21.0941       22.2795
               e  22.4874 0.3657      21.7706      23.2043       21.5475       23.4274
      C        a  18.4840 0.6568      17.1967      19.7712       16.7961       20.1719
               b  19.1254 0.3505      18.4385      19.8123       18.2247       20.0261
               c  19.9303 0.3330      19.2777      20.5830       19.0745       20.7861
               d  20.8959 0.3365      20.2364      21.5555       20.0311       21.7608
               e  22.1366 0.4907      21.1748      23.0984       20.8754       23.3978


Potential outcome mean estimates under A = 1:
---------------------------------------------
 Source Subgroup Estimate     SE Lower 95% CI Upper 95% CI Lower 95% SCB Upper 95% SCB
      A        a  24.3761 0.4201      23.5528      25.1994       23.2965       25.4556
               b  23.6288 0.3126      23.0161      24.2414       22.8254       24.4321
               c  26.7039 0.2630      26.1883      27.2194       26.0278       27.3799
               d  26.6744 0.3081      26.0706      27.2782       25.8827       27.4662
               e  26.7377 0.4499      25.8559      27.6194       25.5814       27.8939
      B        a  27.0322 0.6572      25.7440      28.3203       25.3431       28.7212
               b  26.5918 0.4625      25.6853      27.4983       25.4032       27.7804
               c  29.6127 0.3570      28.9131      30.3123       28.6953       30.5301
               d  29.3991 0.3866      28.6413      30.1569       28.4055       30.3928
               e  28.9304 0.5916      27.7708      30.0899       27.4099       30.4509
      C        a  26.5474 0.9752      24.6360      28.4587       24.0411       29.0537
               b  25.3291 0.5673      24.2173      26.4409       23.8712       26.7869
               c  28.2184 0.5442      27.1518      29.2851       26.8198       29.6171
               d  28.0628 0.5617      26.9619      29.1638       26.6192       29.5064
               e  28.2276 0.7459      26.7656      29.6896       26.3106       30.1447


SuperLearner libraries used:
----------------------------
Outcome model: SL.glmnet, SL.nnet, SL.glm
Treatment model: SL.glmnet, SL.nnet, SL.glm
Source model: NA (model fit via MN.glmnet)
\end{verbatim}
\end{document}